\documentclass[aps,prl,nopacs,twocolumn,preprintnumbers,notitlepage,amsmath,amssymb,superscriptaddress]{revtex4-2}
\usepackage[utf8]{inputenc}
\usepackage{amsmath}
\usepackage{amssymb}
\usepackage{graphicx}
\usepackage{import}
\usepackage{color}
\usepackage{tikz}
\usepackage{makecell}
\usepackage{booktabs}
\usepackage{bm}
\usepackage{placeins}
\usepackage[subnum]{cases}
\usepackage{hyperref}
\usepackage[capitalise]{cleveref} 

\graphicspath{{Images/}{figs/}}

\newcommand{\mup}{\mu^\mathrm{(p)}}
\newcommand{\mus}{\mu^\mathrm{(s)}}
\newcommand{\rhoz}[1]{\rho_{0 #1}}

\newcommand{\selfint}[1]{\alpha_{#1} \Delta\mu_{#1} \rhoz{#1}}
\newcommand{\normselfint}[1]{\frac{\selfint{#1}}{\selfint{1}}}
\newcommand{\prodprod}[1]{\alpha_{#1} \mup_{#1} \rhoz{#1}}
\newcommand{\substprod}[1]{\alpha_{#1} \mus_{#1} \rhoz{#1}}

\begin{document}
\title{Network effects lead to self-organization in metabolic cycles of self-repelling catalysts}
\author{Vincent Ouazan-Reboul}
\affiliation{Max Planck Institute for Dynamics and Self-Organization, Am Fassberg 17, D-37077, G\"{o}ttingen, Germany}
\author{Ramin Golestanian}\email{ramin.golestanian@ds.mpg.de}
\affiliation{Max Planck Institute for Dynamics and Self-Organization, Am Fassberg 17, D-37077, G\"{o}ttingen, Germany}
\affiliation{Rudolf Peierls Centre for Theoretical Physics, University of Oxford, OX1 3PU, Oxford, UK}
\author{Jaime Agudo-Canalejo}\email{jaime.agudo@ds.mpg.de}
\affiliation{Max Planck Institute for Dynamics and Self-Organization, Am Fassberg 17, D-37077, G\"{o}ttingen, Germany}

\begin{abstract}
    Mixtures of particles that interact through phoretic effects are known to aggregate if they belong to species that exhibit attractive self-interactions. We study self-organization in a model metabolic cycle composed of three species of catalytically-active particles that are chemotactic towards the chemicals that define their connectivity network. We find that the self-organization can be controlled by the network properties, as exemplified by a case where a collapse instability is achieved by design for self-repelling species. Our findings highlight a possibility for controlling the intricate functions of metabolic networks by taking advantage of the physics of phoretic active matter.
\end{abstract}

\maketitle

\textit{Introduction.}---\label{sec:intro} Catalyzed chemical reactions are intrinsically and locally out of equilibrium, making catalytic particles a paradigmatic example of systems in which the physics of active matter comes into play \cite{Golestanian2019phoretic}. In particular, catalytic activity coupled to a chemotactic, gradient-response mechanism such as diffusiophoresis \cite{anderson1989Colloid,julicher2009generic} enables the self-propulsion of
individual colloidal particles \emph{via} self-phoresis \cite{paxton2004catalytic,golestanian2007designing}, as well as collective behaviour mediated by effective interactions between active colloids \cite{theurkauff2012dynamic,buttinoni2013dynamical,saha2014clusters,pohl2014dynamic,liebchen2017phoretic}. In addition, catalytic activity is essential to the function of biological systems,
allowing for the occurrence, as a part of metabolism, of reactions that would otherwise be kinetically inhibited \cite{phillips2012Physical}. Metabolic processes often require enzymatic catalysis to occur in a space- and time-localized manner, necessitating some degree of self-organization of the participating enzymes \cite{kerfeld2018bacterial,liu2016cytoophidium,selwood2012dynamic,dueber2009synthetic,kufer2008single,hinzpeter2019regulation,hinzpeter2022trade,Xiong2022}. In particular, many enzymes have been shown to spontaneously form transient aggregates, known as metabolons \cite{sweetlove2018Role}.

Simple cases of spontaneous self-organization in mixtures of several catalytic components have been previously studied both in theory \cite{soto2014self,agudo-canalejo2019Active,grauer2020swarm,giunta2020cross,ouazan-reboul2021Nonequilibrium,cotton2022} and in experiment \cite{yu2018chemical,schmidt2019light,meredith2020predator,testa2021Sustained,grauer2021active}. However, the influence of the sometimes complex topology of reaction networks on the self-organization of the metabolic components has not yet been elucidated. Indeed, many catalytic processes of biological and industrial significance---from cellular metabolism \cite{wu2015krebs} to carbon fixation \cite{schwander2016synthetic}---involve a closed chain of catalytic reactions, where the product of one catalyst is passed on as the substrate of the next one, i.e.~a metabolic cycle. Because the spatial arrangement of catalysts may strongly affect the overall rate of the reactions \cite{hinzpeter2019regulation,hinzpeter2022trade,cotton2022}, it is important to understand under which conditions such spatial reorganization may happen spontaneously, and whether it can be triggered in generically stable systems via network-mediated effects. In this Letter, we study the chemotactic self-organization of three species of catalytically active particles that participate in a model catalytic cycle. We find that a mixture of only self-repelling catalytic species can undergo self-organization via network effects emerging from the metabolic cycle topology.

\begin{figure}[b]
    \begin{center}
        \includegraphics[width=1.0\columnwidth]{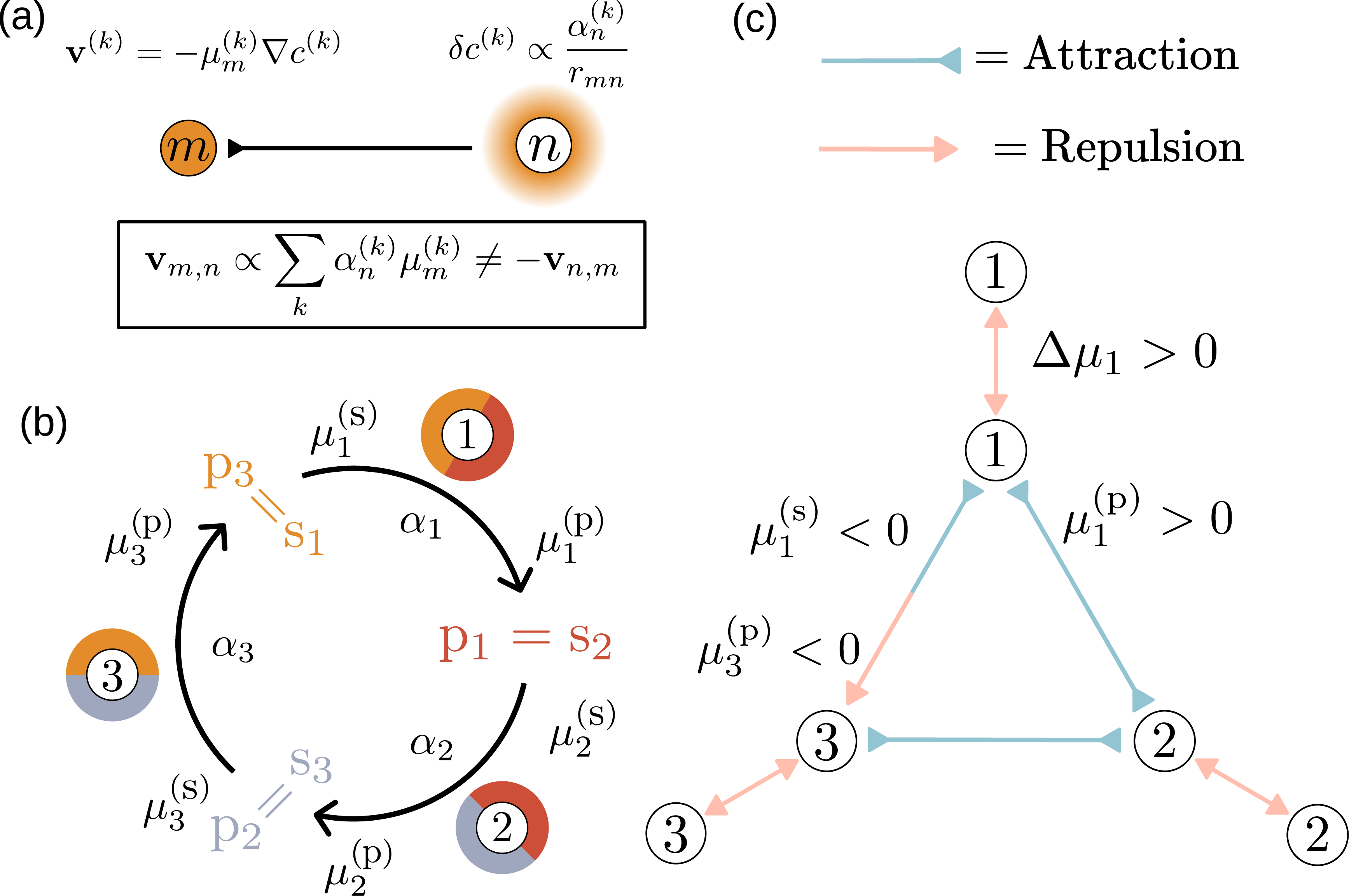}
    \end{center}
    \vskip-0.5cm 
    \caption{ (a) Emergence of field-mediated, nonreciprocal interactions between  particle $n$, which perturbs the chemical field $ c^{(k)} $ around itself, and particle $m$, which develops a velocity in response to the resulting chemical gradient.
    (b) Metabolic cycle of three species. Each converts their substrate into a product which acts as the substrate of the next species, and moves in response to gradients of both chemicals.
    (c) Example of a set of species-species interactions emerging from the combination of effective field-mediated interactions and metabolic cycle topology. With this particular choice of parameters, each species is self-repelling, and the species pair 1-3 exhibits chasing interactions.
    }
    \label{fig:act_mob_cycle}
\end{figure}

\textit{Model.}---\label{sec:interactions}We consider catalytically-active particles
which produce and consume a set of chemicals, with activities $\alpha_m^{(k)}$ corresponding to the rate of production (if positive) or consumption (if negative) of chemical $ k $ by active species $ m $.
These particles are also chemotactic: synthetic colloids typically move \emph{via} hydrodynamic-phoretic mechanisms \cite{Golestanian2019phoretic,yu2018chemical,schmidt2019light,grauer2021active}; whereas the mechanism underlying the observed chemotaxis of biological enzymes is still debated
\cite{yu2009Molecular,sengupta2013Enzyme,dey2014Chemotactic,zhao2018Substratedriven,jee2018Catalytic,agudo-canalejo2018Enhanced,agudo-canalejo2018Phoresis,agudo-canalejo2020Diffusion,feng2020Enhanced}.
In a concentration gradient of chemical $k$, species $m$ develops a velocity $\mathbf{v}_m^{(k)} = - \mu_m^{(k)} \bm{\nabla} c^{(k)}$, where $\mu_m^{(k)}$ is a mobility coefficient. The combination of catalytic and chemotactic activities results in effective interactions between the active species going as $\mathbf{v}_{m,n} \propto \sum_k \alpha_n^{(k)} \mu_m^{(k)}$, where $\mathbf{v}_{m,n}$ represents the velocity of $m$ in response to the presence of $n$; see \cref{fig:act_mob_cycle}(a). Importantly, these interactions are nonreciprocal, in the sense that we generically have  $\mathbf{v}_{m,n} \neq -\mathbf{v}_{m,n}$ \cite{soto2014self}.

We use a continuum model for the concentration $\rho_m$ of the active species $m$ and the chemical concentration $c^{(k)}$, which reads
\begin{subequations}
    \label{eq:evol}
    \begin{equation}
        \label{eq:rho_i_time_ev}
        \partial_t \rho_m (\mathbf{r}, t) 
        =
        \bm{\nabla} \cdot
        \left[
            D_\mathrm{p} \bm{\nabla} \rho_m
            +
            \sum_k \mu_m^{(k)} (\bm{\nabla} c^{(k)}) \rho_m 
        \right],
    \end{equation}\vskip-0.7cm
    \begin{equation}
        \label{eq:c_time_ev}
        \partial_t c^{(k)} (\mathbf{r}, t) 
        = 
        D^{(k)} \bm{\nabla}^2 c^{(k)} 
        + \sum_m \alpha_m^{(k)} \rho_m.
    \end{equation}
\end{subequations}
Here, \cref{eq:rho_i_time_ev} is a continuity equation with $D_\mathrm{p}$ being the diffusion coefficient of the colloidal particles, which we assumed to be equal for all active particles for simplicity, and the drift velocity following the concentration gradients of all chemicals. Moreover, \cref{eq:c_time_ev} is a reaction-diffusion equation for the chemicals, with $D^{(k)}$ representing the diffusion coefficient of chemical $(k)$, and the reaction term accounting for the local activity of all catalytic species.

To determine when such a mixture undergoes an instability, we perform a linear stability analysis of these equations \cite{suppmat}. We find that a perturbation $( \delta \rho _m, \delta c ^{(k)})$ around an initially homogeneous state $(\rhoz{m}, c_h^{(k)})$ follows the general eigenvalue equation $-\sum_{n=1}^{M} \Lambda_{m,n} \delta \rho_n = \left[\lambda+ D_{\rm p} q^2\right]\delta \rho_m$
where $ \Lambda_{m,n} = \sum_k \frac{\alpha_n^{(k)} \mu_m^{(k)}}{D^{(k)}} \rhoz{m} $
represents the response of species $m$ to species $n$, mediated by all chemical fields, and $ \lambda (q)$ is the growth rate of a perturbation with wave number $q$. The coefficient $\Lambda_{m,n}$ is negative (positive) if species $m$ is attracted to (repelled by) species $n$. Throughout the rest of this Letter, we rescale the mobility coefficients for brevity, such that $\mu_m^{(k)}/D^{(k)} \to \mu_m^{(k)}$. The system undergoes an instability if any mode has a positive growth rate $ \lambda > 0 $. We focus on the eigenvalue with the highest growth rate associated with the $q^2 = 0$ mode, corresponding to the longest wavelength, which represents a macroscopic instability. 

\textit{Metabolic cycles.}---\label{sec:cat_cycle_form}We focus on the particular case of  metabolic cycles composed of 3 active species (\cref{fig:act_mob_cycle}(b)), where species $m$ converts its substrate ${\rm s}_m$ into its product ${\rm p}_m={\rm s}_{m+1}$, with an activity $\alpha_m = \alpha_m^{(m+1)} = - \alpha_m^{(m)} > 0$, as depicted in \cref{fig:act_mob_cycle}(b).
As the cycle is closed, the species indices are periodic, with species 4 being identical to species 1, and species 0 to species 3.

The species have a chemotactic response to both their substrate and product, with respective mobilities $\mus_m \equiv \mu_m^{(m)} $ and $\mup_m \equiv \mu_m^{(m+1)}$.
The resulting interaction matrix has the following (non-vanishing) coefficients (\cref{fig:act_mob_cycle}(c)): 
$\Lambda_{m,m-1}=\alpha_{m-1}\mus_m\rhoz{m}$,  $\Lambda_{m,m}=\alpha_m \Delta \mu_m \rhoz{m}$,  and $\Lambda_{m,m+1}=-\alpha_{m+1}\mup_m\rhoz{m}$, where $ \Delta \mu_m \equiv \mup_m - \mus_m $, and is in general asymmetric, reflecting the non-reciprocal nature of the interactions between the catalytic species.

We now calculate the eigenvalues
$\lambda(q=0)$ for $\Lambda_{m,n}$ as defined above, and find two eigenvalues given by \cite{suppmat}
\begin{widetext}
    \begin{equation}
        \label{eq:evals}
            \lambda_\pm 
            =
            - \frac{1}{2}
            \sum\limits_{m=1}^3 
                \alpha_m \Delta \mu_m \rhoz{m}
            \pm
            \frac{1}{2}
            \sqrt{
                \left( 
                    \sum\limits_{m=1}^3 
                    \alpha_m \Delta \mu_m \rhoz{m} 
                \right)^2
                -  
                4 \sum\limits_{m=1}^3 
                    \alpha_m \alpha_{m+1}
                    \left( 
                        \Delta \mu_m \Delta \mu_{m+1}
                        +
                         \mup_{m} \mus_{m+1}
                    \right)
                    \rhoz{m} \rhoz{m+1} 
            },
    \end{equation}
\end{widetext}
as well as one null eigenvalue $ \lambda_0 = 0 $. The system will be linearly unstable when (the real part of) the largest eigenvalue $ \lambda_+ $ becomes positive.

\textit{Substrate-sensitive species.}---\label{sec:phase_diag_3_1mob}A simple class of cycles whose parameter space can be explored in full generality involves those in which the catalytic particles are only chemotactic towards their substrate, i.e.~$\mup_m =0$. We thus have three activities $\alpha_m$ and three substrate mobilities $\mus_m$, with the mobility difference reducing to $ \Delta \mu_m  = - \mus_m$. Species in such a cycle then only interact with the previous species in the cycle and with themselves, with $\Lambda_{m,m-1} = \alpha_{m-1} \mus_m \rhoz{m}$, $\Lambda_{m,m} = -\alpha_{m} \mus_m \rhoz{m}$, and $\Lambda_{m,m+1} = 0$. Note that the self-interaction always has the opposite sign to the interaction with the previous species in the cycle, which further limits the possible interaction patterns the catalytic species can exhibit.

\begin{figure}
    \centering
    \includegraphics[width = \columnwidth]{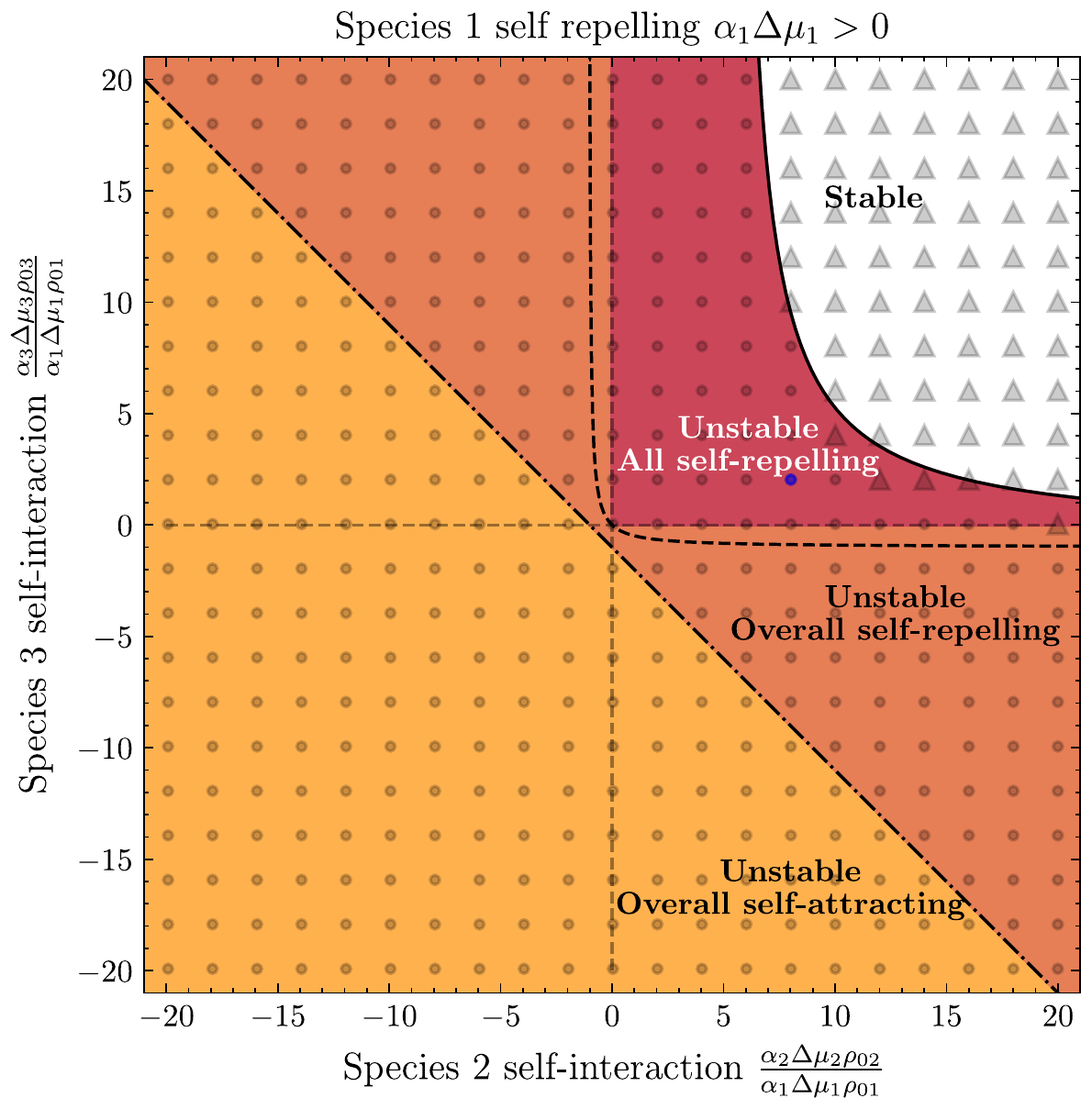}
    \caption{
        {Stability phase diagram for cycles involving at least one self-repelling species.}
        Species 1 taken to be self-repelling ($ \alpha_1 \Delta \mu_1 > 0 $).
        Full line: Stability line in normalized self-interaction plane for a specific choice of parameters.
        Dashed line: Parameter-free instability line below which cycles with null product mobilities $ \mup_{1,2,3} = 0
        $ are unstable. Note that this line always below the the top-right, all-self-repelling quadrant.
        Dash-dotted line: boundary between an overall self-attracting and self-repelling catalytic mixtures.
        Gray markers: coordinates of Brownian dynamics simulations, found to be unstable if the marker is a circle or stable if it is a triangle.
        The blue marker corresponds to the coordinates of the simulation shown in \cref{fig:3_cycle_snaps}.
        For the expressions of the stability lines and the values of the parameters, see the Supplemental Material
        \cite{suppmat}.
    }
    \label{fig:omob_ph_diag}
\end{figure}

In the context of this reduced parameter space, and assuming species 1 is self-repelling (i.e. $ \alpha_1 \Delta \mu_1 > 0 $), one can solve $
\operatorname{Re} (\lambda_+) > 0 $ for all parameter
values, yielding a comprehensive two-dimensional stability phase diagram as shown in \cref{fig:omob_ph_diag} \cite{suppmat}, which depends only the normalized self-interactions 
\begin{math}\displaystyle
\normselfint{2}
\end{math}
and 
\begin{math}\displaystyle
\normselfint{3}.
\end{math}
The corresponding parameter-free instability line is plotted as a dashed line on \cref{fig:omob_ph_diag}, with the dark orange and light orange regions below that line corresponding to unstable metabolic cycles. As the instability line is above the light orange region corresponding to overall self-attracting mixtures in \cref{fig:omob_ph_diag} ($ \sum_{m=1}^3 \Lambda_{m,m} < 0 $), we uncover that it not necessary for the metabolic cycle to be composed of overall self-attracting species in order to self-organize, as opposed to mixtures involving simpler interaction schemes \cite{agudo-canalejo2019Active,ouazan-reboul2021Nonequilibrium,ouazanreboul2022preprint}. This result does not, however, extend to cycles composed only of self-repelling species ($\Lambda_{m,m} > 0 $ for all $m$), which are always stable, as shown by the fact that the corresponding top right quadrant in \cref{fig:omob_ph_diag} is always above the dashed stability line. This implies that some amount of self-attraction is still necessary for the catalytic particles to self-organize in this limited interaction topology.

\textit{Self-organization of purely self-repelling species}---\label{sec:cycle_decomp}We now consider the general case
with both nonzero substrate and product mobilities, for which each species interacts with both the previous and the next species in the cycle according to a pattern set by its substrate mobility $ \mus $, its product mobility $ \mup $, and their difference $ \Delta \mu$.
By solving for $ \operatorname{Re} (\lambda_+) > 0 $ in \cref{eq:evals}, we find that a cycle which is overall self-repelling can be made unstable provided the following condition is satisfied
\begin{equation}
    \label{eq:gen_instab_2}
    \begin{aligned}
        \sum\limits_{m=1}^{3} &\alpha_m \Delta\mu_m \rhoz{m} \cdot \alpha_{m+1} \Delta\mu_{m+1} \rhoz{m+1}\\
                              & < - \sum\limits_{m=1}^{3} \alpha_m\mus_{m+1}\rhoz{m+1} \cdot \alpha_{m+1}\mup_{m}\rhoz{m}.
    \end{aligned}
\end{equation}
The condition (\cref{eq:gen_instab_2}), which involves terms mixing pairs of catalytic species, can be rewritten as $\sum \Lambda_{m,m} \Lambda_{m+1,m+1} < \sum \Lambda_{m,m+1}\Lambda_{m+1,m}$, and thus sets a bound on the self-interactions of pairs of species relative to their cross-interactions. This inequality implies that, in the case of an overall self-repelling cycle where $ \sum_{m=1}^3 \Lambda_{m,m} > 0 $, the presence of species pairs which interact reciprocally offer an alternate route to instability.

We find that a striking new feature of this general case is that cycles composed only of self-repelling species can be unstable according to the condition given in \cref{eq:gen_instab_2} if any self-repelling species $ m $ verifies one of the following inequalities
\cite{suppmat}:
\begin{subequations}
    \begin{align}
        & \hskip1.cm \prodprod{m+1} > \selfint{m}, \label{cond1}  \\ \nonumber \\
        & \hskip3.4cm \mus_m > 0, \label{cond2}  \\
        & \frac{1}{ \alpha_m \mus_m \rhoz{m} }\Big[\prodprod{m-1}  \substprod{m} \nonumber \\
        &\hskip1.5cm +\prodprod{m+1} \Big( \prodprod{m} \nonumber \\
        &\hskip2.5cm + \prodprod{m-1} \Big) \Big] > 0 . \label{cond3}
    \end{align}
\end{subequations}
This behavior is illustrated, for a particular set of parameters for which species 1 obeys inequalities \cref{cond1,cond3}, by the stability phase diagram shown in \cref{fig:omob_ph_diag}.
For this choice of parameters, the stability line is contained in the top-right quadrant, which corresponds to a mixture of three self-repelling species. 
Thus, in some regions of the parameter space (\cref{fig:omob_ph_diag}, dark red), the destabilizing pair interactions are able to overcome self-repulsion, and lead to the formation of a cluster of three species in the absence of self-attraction. We recall that cycles of strictly substrate-sensitive species, which are stable above the dashed line drawn in \cref{fig:omob_ph_diag}, exhibit no such region independently of the choice of parameters. An increased complexity of interactions is therefore needed in order to overcome self-repulsion.
We have confirmed this analytical prediction using Brownian dynamics simulations (\cref{fig:omob_ph_diag}, grey circles and triangles). By scanning the coordinates of \cref{fig:omob_ph_diag} for a set of parameters given in \cite{suppmat}, we broadly
recover the predicted instability line.

\begin{figure*}[hbtp]
    \centering
    \includegraphics[width=0.99\textwidth]{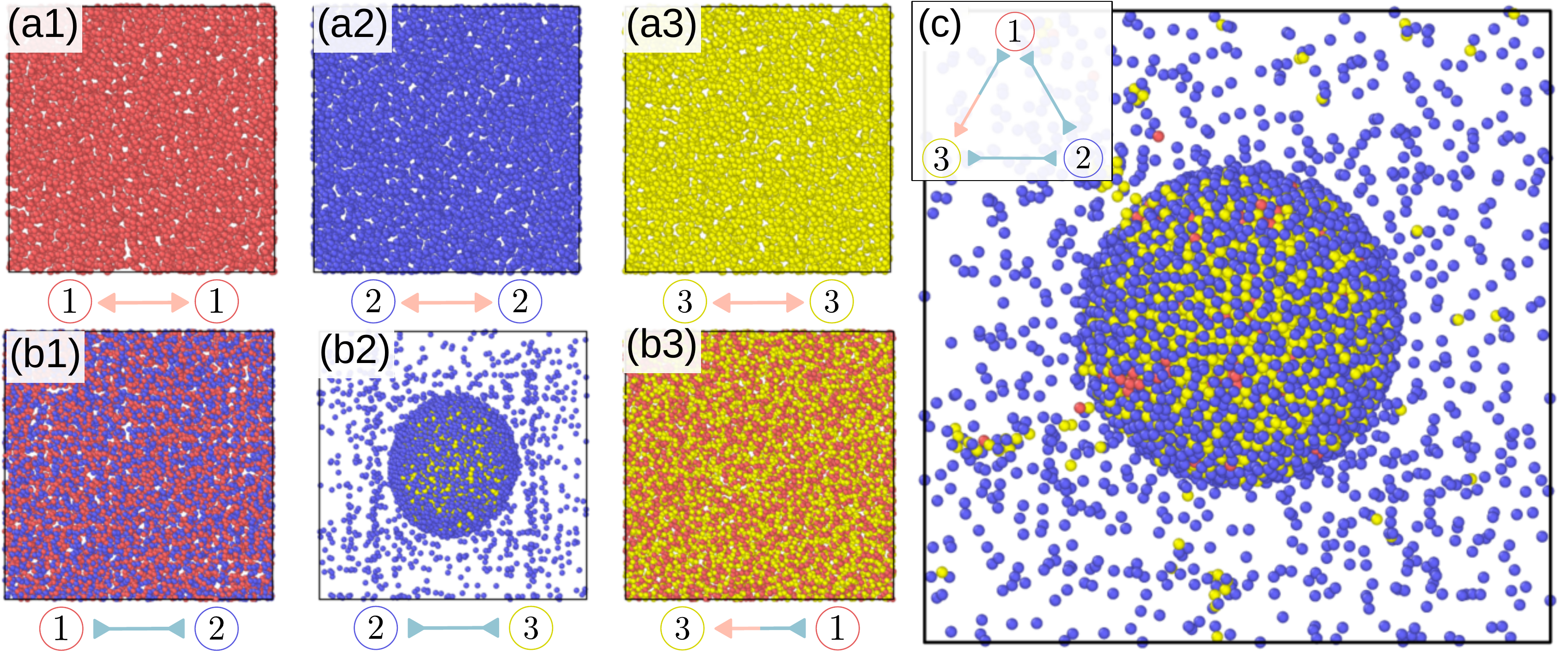}
    \caption{
        {Instability of a generic three-enzyme cycle triggered by the presence of a single instability-favouring pair.}
        (a1--a3) All particle species are self-repelling, and thus single-species suspensions stay homogeneous.
        (b1--b3) Of the two-species mixtures, 1-2 and 3-1 mixtures form small molecules but remain homogeneous, while 2-3 is unstable and results in the formation of a 2-3 cluster coexisting with a gas of species 2. 
        (c) Despite all species being self-repelling, the instability of the 2-3 pair causes the instability of the full three-species mixture resulting in the formation of a cluster, which coexists with a gas of species 2.
        Note that species 1 (red) also participates in the cluster,  despite not aggregating with either of the other species (b1, b3).
        See also Movies 1--7 and the Supplemental Material \cite{suppmat} for a description of the simulation parameters and movies.
    }
    \label{fig:3_cycle_snaps}
\end{figure*}
The results of a simulation of an unstable cycle of three self-repelling species, with parameters corresponding to the blue point in \cref{fig:omob_ph_diag} and the interactions shown in \cref{fig:act_mob_cycle}(c), are shown in \cref{fig:3_cycle_snaps} (see \cite{suppmat} for simulation parameters). In panels (a1) to (a3), we find that each catalytic species taken on its own does not cluster, because of being self-repelling. Panels (b1) and (b3) demonstrate that mixtures of particles belonging to species pairs (1,2) and (3,1), when considered together, are also stable, and lead to the formation of small dynamic molecules. However, as shown in panel (b2), mixtures of particles of species 2 and 3 are unstable, and lead to the formation of a mixed clusters coexisting with a dilute phase. A favourable choice of parameters allows species 2 and 3 to destabilize the whole ternary (1,2,3) mixture of catalysts, despite the stabilizing effects of the individual species and the rest of the species pairs. The resulting structure is a cluster mixing all three species coexisting with a dilute phase.

\textit{Discussion.}---Using a simple model, we have shown that three catalytically active species involved in a model metabolic cycle are able to undergo a self-organizing instability through chemical field-mediated effective interactions. In the case of a cycle involving species which are only chemotactic towards their substrates, we find that while self-organization is possible if the mixture of catalytic species is overall self-repelling, at least one of the species must be self-attracting for the instability to occur. However, in contrast to this case and to what was previously known for phoretic particles
\cite{agudo-canalejo2019Active,ouazan-reboul2021Nonequilibrium,ouazanreboul2022preprint}, we find that the system can self-organize even when all species are self-repulsive through instability-favouring pair interactions, in the general case of species chemotactic to both their substrates and products.

We found here that the topology of interactions between catalytically active particles can lead to entirely new forms of collective behaviour, namely self-organization of self-repelling particles. While we considered minimal catalytic cycles with only three species, our work also shows that the number of species in the cycle can have a strong effect on the resulting dynamics, with cycles of even and odd number of species displaying entirely different behaviour \cite{ouazanreboul2022preprint}. Future work may consider more complex and biologically relevant topologies of the catalytic network (e.g.~branched), as well as the effect of self-organization on the metabolic properties of the system \cite{dueber2009synthetic,hinzpeter2019regulation,hinzpeter2022trade,Xiong2022}. Furthermore, our work may find application in engineering synthetic functional structures with shape-shifting capacity at the molecular scale \cite{Osat2022}.

\acknowledgments
This work has received support from the Max Planck School Matter to Life and the MaxSynBio Consortium, which are jointly funded by the Federal Ministry of Education and Research (BMBF) of Germany, and the Max Planck Society.

\end{document}


\title{
    Network effects lead to self-organization in metabolic cycles of self-repelling catalysts
    \\
    \textit{Supplemental Material}
    }
    
\author{Vincent Ouazan-Reboul}
\affiliation{Max Planck Institute for Dynamics and Self-Organization, Am Fassberg 17, D-37077, G\"{o}ttingen, Germany}
\author{Ramin Golestanian}\email{ramin.golestanian@ds.mpg.de}
\affiliation{Max Planck Institute for Dynamics and Self-Organization, Am Fassberg 17, D-37077, G\"{o}ttingen, Germany}
\affiliation{Rudolf Peierls Centre for Theoretical Physics, University of Oxford, OX1 3PU, Oxford, UK}
\author{Jaime Agudo-Canalejo}\email{jaime.agudo@ds.mpg.de}
\affiliation{Max Planck Institute for Dynamics and Self-Organization, Am Fassberg 17, D-37077, G\"{o}ttingen, Germany}

\maketitle

\tableofcontents

\setcounter{equation}{0}
\setcounter{figure}{0}
\renewcommand{\theequation}{S\arabic{equation}}
\renewcommand{\thefigure}{S\arabic{figure}}
\renewcommand{\thetable}{S\Roman{table}}

\section{Linear stability analysis for three-species cycles}

To calculate the eigenvalues of the matrix $\Lambda_{m,n}$ in the main text, we construct the associated characteristic polynomial, which reads
\begin{equation}
    \chi_{-\Lambda}( \lambda )
    =
    \begin{vmatrix}
        - \alpha_1 \Delta \mu_1 \rhoz{1} - \lambda &  \alpha_2 \mup_1 \rhoz{1} & - \alpha_3 \mus_1 \rhoz{1} \\
        - \alpha_1 \mus_2 \rhoz{2} & - \alpha_2 \Delta \mu_2 \rhoz{2} - \lambda &  \alpha_3 \mup_2 \rhoz{2}   \\
         \alpha_1 \mup_3 \rhoz{3} & - \alpha_2 \mus_3 \rhoz{3} & - \alpha_3 \Delta \mu_3 \rhoz{3} - \lambda    \\
    \end{vmatrix}.
\end{equation}
Performing the operations $ C_1 \leftarrow C_1 + \frac{\alpha_1}{\alpha_2} C_2 + \frac{\alpha_1}{\alpha_3} C_2$ on the columns $C_i$ of the matrix, we obtain
\begin{equation}
    \chi_{-\Lambda}( \lambda )
    =
    \begin{vmatrix}
        - \lambda &  \alpha_2 \mup_1 \rhoz{1} &  - \alpha_3 \mus_1 \rhoz{1} \\
        - \frac{\alpha_1}{\alpha_2} \lambda & - \alpha_2 \Delta \mu_2 \rhoz{2} - \lambda & \alpha_3 \mup_2 \rhoz{2}   \\
        - \frac{\alpha_1}{\alpha_3} \lambda & - \alpha_2 \mus_3 \rhoz{3} & - \alpha_3 \Delta \mu_3 \rhoz{3} - \lambda    \\
    \end{vmatrix},
\end{equation}
which can be further manipulated to give
the following polynomial
\begin{equation}
    \begin{split}
    \chi_{-\Lambda}(\lambda) = 
    -\lambda
    \Big[
        &\lambda^2
        +  
        \left(
            \alpha_1 \Delta\mu_1 + 
            \alpha_2 \Delta\mu_2 + 
            \alpha_3 \Delta\mu_3
        \right)
        \lambda
        \\
        + &\alpha_1\alpha_2
        \left(
            \Delta\mu_1 \Delta\mu_2 + \mup_{1}\mus_{2}
        \right)
        + \alpha_2\alpha_3
        \left(
            \Delta\mu_2 \Delta\mu_3 + \mup_{2}\mus_{3}
        \right)
        + \alpha_3\alpha_1
        \left(
            \Delta\mu_3 \Delta\mu_1 + \mup_{3}\mus_{1}
        \right)
        \Big].
    \end{split}
\end{equation}
We can now solve $\chi_{-\Lambda}(\lambda) = 0$ to obtain Eq. (2) in the main text and $\lambda = 0$.

\section{Stability condition for substrate-sensitive species}
\label{sec:omob}

Here we consider the case of a cycle with three species, with activities $\alpha_m$, substrate mobilities $\mus_m$, and no product mobilities. The corresponding interaction matrix has the two non-null eigenvalues:
\begin{equation}
    \label{eq:omob_eig_reduced}
    \lambda_\pm =
    - \frac{1}{2} \sum_{m=1}^3 \selfint{m}
    \pm \frac{1}{2}
    \sqrt{
        \left( \sum_{m=1}^3 \selfint{m} \right)^2 -
        4   \sum_{m=1}^3 \selfint{m} \cdot \selfint{m+1} 
    },
\end{equation}
with $ \Delta \mu_m = - \mus_m $. In the particular case for which $ \alpha_1 \Delta \mu_1 > 0 $, solving $ \lambda_+ > 0 $ leads to two possible parameter-free instability conditions, which can be written as
\begin{equation}
    \label{eq:omob_type_I}
    \normselfint{3} < - \left( \normselfint{2} + 1 \right)
\end{equation}
and
\begin{equation}
    \label{eq:omob_type_II}
    \normselfint{3} \left( { 1 + \normselfint{2} } \right) < -{ \normselfint{2} },
\end{equation}
with \cref{eq:omob_type_I} corresponding to an overall self-attracting mixture and \cref{eq:omob_type_II} corresponding to a self-repelling one.
We can show that, if 
\begin{math}\displaystyle
 \normselfint{2} < -1,
\end{math}
then the conditions given in \cref{eq:omob_type_I} and \cref{eq:omob_type_II}
are complementary, and the cycle is always unstable.
If, however, 
\begin{math}\displaystyle
 \normselfint{2} \geq -1,
\end{math}
then \cref{eq:omob_type_II} is the least strict stability condition and thus determines 
the overall stability of the cycle. This boundary is plotted as the dashed line in main text Fig. 2, with cycles having parameters below that line
being unstable. It can be shown that the condition given in \cref{eq:omob_type_II} is not achievable for a cycle of three self-repelling species.

\section{Self-repelling metabolic cycles} 
\label{sec:generic}
We use a similar approach to \cref{sec:omob} in order to find the instability conditions for a cycle of three self-repelling species with non-null product mobilities $ \mup $. We seek to solve $ \operatorname{Re}(\lambda_+) > 0 $ (see main text Eq. (3)), under the assumption that the mixture of catalysts is self-repelling, i.e. the following condition holds $ \sum_{m=1}^{3} \selfint{m} > 0$.

In order to obtain expressions in the normalized self-interaction plane 
\begin{math}\displaystyle
\left( \normselfint{2}, \normselfint{3}
\right), 
\end{math}
we rewrite main text Eq. (3) in terms of the $ \Delta \mu_m $ and $ \mup_m $.
For brevity, we introduce the notations 
\begin{math}\displaystyle
\Delta_m \equiv \normselfint{m} 
\end{math}
and 
\begin{math}\displaystyle
p_m \equiv \frac{\prodprod{m}}{\selfint{1}}.
\end{math}
This results in the following form for the instability condition
\begin{equation}
    \label{eq:instab_gen}
    \left( \Delta_2 - p_2 + 1 \right) \Delta_3
    <
    \left(  p_1 - 1 \right) \Delta_2
    -  \sum_{m=1}^{3} p_m p_{m+1} 
    + p_3,
\end{equation}
which we can plot in normalized self-interaction coordinates $ \Delta_2, \Delta_3 $ by taking $ p_1, p_2, p_3 $ as arbitrary parameters. Defining
\begin{equation}
    \label{eq:instab_function}
    f_{\{p_m\}}(\Delta_2) \equiv \frac{ ( p_1 - 1 )  \Delta_2 - \sum_{m=1}^{3} p_m p_{m+1} + p_3 }{ \Delta_2 - p_2 + 1 },
\end{equation}
and rewriting \cref{eq:instab_gen} as 
\begin{equation}
    \label{eq:instab_reduced}
    \begin{cases}
        \Delta_3 < f_{\{p_m\}}(\Delta_2) \text{ if } \Delta_2 > p_2 - 1, \\
        \Delta_3 > f_{\{p_m\}}(\Delta_2) \text{ if } \Delta_2 \leq p_2 - 1 ,
    \end{cases}
\end{equation}
\\
we can then determine the conditions for which \cref{eq:instab_reduced} can be satisfied if all three species are self-repelling,
i.e. $ \Delta_2 > 0 \text{ and } \Delta_3 > 0 $.

This statement is equivalent to having at least part of the instability line located in the $ \Delta_2 > 0, \Delta_3 > 0 $ domain of the reduced self-interaction plane in which we plot the instability line. Thus, if there is at least one $ \Delta_2 > 0 $ such that $ f_{\{p_m\}}(\Delta_2) > 0 $, an instability can be obtained for a set
of three self-repelling species.

The function $ f_{\{p_m\}} $ features two asymptotes, a vertical one with value $ p_2 - 1 $ and a horizontal one with value $ p_1
- 1$. If either of these asymptotes are positive, then the existence of a region of $ \Delta_2 > 0 $ such that $
f_{\{p_m\}}(\Delta_2) > 0  $ is guaranteed, leading to the first two conditions written in main text Eqs. (4a) and (4b).
If the two asymptotes are negative, there is still a possibility for all-self-repelling instabilities, as $ f_{\{p_m\}} $ features
a zero with value $ \Delta_2^{(0)} =  (\sum_{m=1}^{3} p_m p_{m+1} - p_3) / (p_1 - 1) $.
If $ \Delta_2^{(0)} > 0 $, $ f $ exhibits a sign change in the right half-plane and is guaranteed to be positive-valued
in parts of it.
We rewrite the $ \Delta_2^{(0)} > 0 $ inequality in terms of activities, mobilities, and homogeneous concentrations, and
present a compact form of it in main text Eq. (4c).
As main text Eq.~(3) is invariant by cyclic permutation of the catalytic species, the obtained conditions generalize to any of the three species
index $ m \in \left\{ 1, 2, 3 \right\} $, as long as the corresponding species is self-repelling.
In the conditions written in main text Eq.~(4a-c), we then replace species 1 by an arbitrary species $m$, species 2 with species $m+1$, and species 3
with $m-1$.

In main text Fig. 2, we plot the stability line corresponding to $ p_1 = 0.473 $, $ p_2 = 6.315 $, and $ p_3 = -5.684 $.
As $ p_1 < 1 $ and $ p_2 > 1 $, the horizontal and vertical asymptotes of the stability function $ f $ are, respectively, negative and positive.
Furthermore, as the corresponding value of $ \Delta_2^{(0)} $ is positive, the stability function exhibits a sign change in the right half-plane.
Note that we only plot the stability line for $ \Delta_2 > p_2 - 1 $, as the values of $ f_{\{p_m\}} $ for $ \Delta_2 \leq p_2 - 1 $ are systematically below the overall-self-attracting line and thus irrelevant.

\begin{figure}
    \centering
    \includegraphics[width=0.5\textwidth]{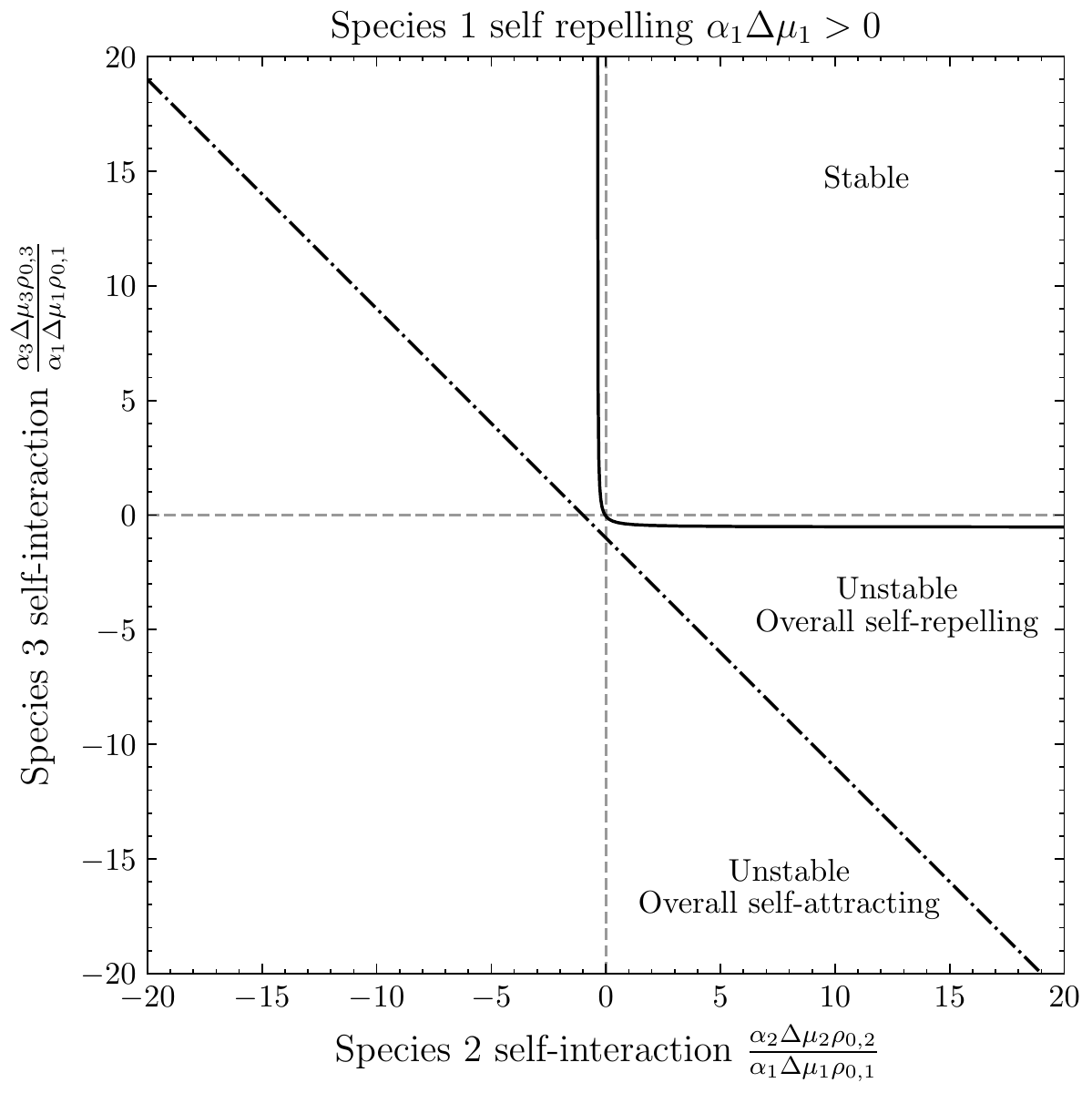}
    \caption{ Phase diagram with parameters $ \divbyone{ \alpha_1 \mup_1 \rhoz{1} } = 0.473 $, $ \divbyone{ \alpha_2
        \mup_2 \rhoz{2} } = 0.6315 $, and $ \divbyone{ \alpha_3 \mup_3 \rhoz{3} } =
        -2.684 $, without an unstable region for all-self-repelling catalytic species.
    }
    \label{fig:no_unst_full_sr}
\end{figure}

On the other hand, if we choose for instance $ p_1 = 0.473 $, $ p_2 = 0.6315 $, and $ p_3 = -2.684 $, both asymptotes
are negative and the sign change of $ f_{\{p_m\}} $ occurs in the right half-plane, as shown in \cref{fig:no_unst_full_sr}.

\section{Brownian dynamics simulations}

To study the system beyond the linear regime,  we perform Brownian Dynamics simulations.
We simulate the dynamics of chemically active and chemotactic spherical particles, the effective interactions among
which we derive in the following.
We first consider an isolated particle of species $m$,  with activity $\alpha_m$ and radius $R$.
We place the particle at the origin, and use spherical coordinates.  
The perturbation $\delta c^{(k)}_m$ induced by the particle to chemical $k$,  which is assumed to equilibrate quickly with respect to the motion of the particle,  is given by the solution of the Laplace equation:
\begin{equation}
    \label{eq:evol_c_MD}
    0 = 
    \nabla^2 \delta c^{(k)}_m.
\end{equation}
The corresponding boundary conditions, however,   depend on whether the chemical is the substrate of species $m$ (i.e. $k=m$), its product (i.e. $k=m+1$), or neither.
Indeed, the boundary condition must balance the diffusive fluxes across the particle surface with its chemical activity,
resulting in:
\begin{equation}
    \label{eq:evol_c_MD_BC}
    -4 \pi R^2 D^{(k)}
    \left. \frac{\partial \delta c^{(k)}_m}{\partial r} \right|_{r=R}  = 
    \begin{cases}
    -\alpha_m & \text{if $k=m$}, \\
    \alpha_m & \text{if $k=m+1$}, \\
    0 & \text{otherwise}.
    \end{cases}
\end{equation}
The corresponding solutions for the perturbations are then
\begin{equation}
    \label{eq:c_sph_part}
    \delta c^{(k)}_m (r) = 
    \begin{cases}
    - \frac{\alpha_m}{4 \pi D^{(k)}} \frac{1}{r}  & \text{if $k=m$},  \\
    \frac{\alpha_m}{4 \pi D^{(k)}} \frac{1}{r} & \text{if $k=m+1$}, \\
    0 & \text{otherwise}.
    \end{cases}
\end{equation}

Now consider a second particle of species $n$ placed at a location $\mathbf{r}$.
Its velocity $\mathbf{v}_{n,m}(\mathbf{r})$ in response to the perturbation(s) created by the particle of species $m$ will be:
\begin{equation}
    \label{eq:MD_resp_base}
    \mathbf{v}_{n,m} (\mathbf{r})  = 
    \begin{cases}
    - \mus_{n} \nabla \delta c^{(m+1)}_m & \text{if $n=m+1$}, \\
    - \mup_{n} \nabla \delta c^{(m)}_m & \text{if $n=m-1$}, \\
    - \mus_{n} \nabla \delta c^{(m)}_m-\mup_{n} \nabla \delta c^{(m+1)}_m & \text{if $n=m$}, \\
    0 & \text{otherwise}.
    \end{cases}
\end{equation}
Using Eq.~\eqref{eq:c_sph_part}, the responses can be explicitly written as:
\begin{equation}
    \label{eq:full_MD_response}
    \mathbf{v}_{n,m}(\mathbf{r})  = 
    \begin{cases}
    \frac{\alpha_{m}}{4 \pi} \frac{\mus_{n}}{D^{(n)}} \frac{\mathbf{r}}{{r}^3} & \text{if $n=m+1$}, \\
    -\frac{\alpha_{m}}{4 \pi} \frac{\mup_{n}}{D^{(n+1)}} \frac{\mathbf{r}}{{r}^3} & \text{if $n=m-1$}, \\
    \frac{\alpha_{m}}{4 \pi} \left( \frac{\mup_{n}}{D^{(n+1)}} - \frac{\mus_{n}}{D^{(n)}} \right) \frac{\mathbf{r}}{{r}^3}  & \text{if $n=m$}, \\
    0 & \text{otherwise}.
    \end{cases}
\end{equation}

We can then simulate the following equations of motion for $N$ particles:
\begin{equation}
    \label{eq:update}
    \dot{\mathbf{r}}_i(t) = 
    \sum_{\substack{j=1 \\ (j\neq i)}}^N \mathbf{v}_{S(i),S(j)}(\mathbf{r}_i-\mathbf{r}_j) + \boldsymbol{\xi}_m
\end{equation}
with $i \in \{1, 2, \ldots, N\}$, $S(i)$ giving the species index corresponding to the particle index $i$, the velocities calculated using Eq.~\eqref{eq:full_MD_response},
and $\boldsymbol{\xi}$ a random velocity corresponding to white noise with intensity $2D_\mathrm{p}$,
resulting in diffusion of the particles with diffusion coefficient $D_\mathrm{p}$.
Particles start out randomly distributed in space, corresponding to a homogeneous state.
The equations of motion are integrated using a forward Euler scheme.
At every integration step, an overlap correction is then performed to account for hard-core repulsion between the
spheres, using the ``elastic collision method'' \cite{strating1999Brownian} and a cell-list algorithm \cite{allen2017Computer} with cubic cells of side length $2 R$  to determine particles' neighbours.
We simulate the system in a three-dimensional box of side length $L$ with periodic boundary conditions, and interactions are treated according to the minimum image convention.
The particle diameter, $\sigma=2 R$, which is taken to be the same for all species, sets the basic length scale of the simulation.
We can redefine the mobility coefficients to include the diffusion coefficient of the corresponding chemical, as $\hat{\mu}^{({\rm s},{\rm p})}_m = \mu^{({\rm s},{\rm p})}_m / D^{(m,m+1)}$.
We can define basic activity and mobility scales, respectively $\alpha_0$ and $\hat{\mu}_0$, from which we build a velocity scale
$V_0 = \alpha_0 \hat{\mu}_0 / ( 4 \pi \sigma^2 ) $.
From these scales, we can define dimensionless time $\tau = t V_0 / \sigma$, activities $\tilde{\alpha} = \alpha / \alpha_0$,
and mobilities $\tilde{\mu} = \hat{\mu} / \hat{\mu}_0$.
Finally, we define a reduced particle diffusion coefficient $\tilde{D}_\mathrm{p} = D_\mathrm{p} / (V_0 \sigma) $, equivalent to a noise intensity or temperature.

We numerically verify the stability  diagrams shown in Fig.~2 of the main text.
To do so, we perform simulations on a twenty by twenty grid in the 
\begin{math}\displaystyle
\normselfint{2}, \normselfint{3}
\end{math}
plane, which we generate with the
parameters given in \cref{tab:scan_params}.
\begin{table}[t]
\begin{tabular}{cccc}
Species m & Activity $ \tilde{\alpha}_m $ & Substrate mobility $ \tmus_m $                                            & Product mobility $ \tmup_m $ \\ \midrule 
1         & 0.125                         & -0.5                                                                      & 0.45 \\
2         & 0.75                          & $ -2.166 + 0.316 \cdot n_2, \ n_2 \in \left\{ 1, 2, \ldots, 20 \right\} $ & 1 \\
3         & 0.75                          & $ -4.066 + 0.316 \cdot n_3, \ n_3 \in \left\{ 1, 2, \ldots, 20 \right\} $ & -0.9\\ \bottomrule
\end{tabular}
\caption{Simulation parameters for main text Fig. 2.}
\label{tab:scan_params}
\end{table}

We then perform Brownian Dynamics simulations for all possible combinations of $n_2 \text{ and } n_3 \in \left\{
1, 2 \ldots, 20 \right\}$, making for a total of 441 sets of parameters.
5000 particles of each species are simulated, in a box of size $ L/\sigma = 54 $, corresponding to a total volume fraction $
\phi = 0.05 $.
The simulations are iterated with a time step $ \mathrm{d}\tau = 0.005 $, for a total time of $ \tau_\text{total} = 60
$.
The effective diffusion coefficient is set at $ \tilde{D}_\mathrm{p} = 0.01 $.
The simulations are run in the Julia programming language.
To determine whether the system is unstable, we perform a clustering analysis on the final state of the system using a
cell-lists based clustering algorithm \cite{allen2017Computer}, and considering that two particles are neighbours if they are
within a distance of $ 1.1 \sigma $.
A simulation run is considered to be unstable if, at the end of the runtime, any cluster includes more than 20\% of the
total particle population.
We obtain the results shown in Fig.~2 of the main text.
Overall, the numerical results agree with the analytical prediction for the instability line.
\\
The results shown in Fig.~3 of the main text correspond to the steady states of simulations ran with the parameters
compiled in  \cref{tab:scan_params} with $ n_2 = 6 \text{ and } n_3 = 9 $ (i.e. $ \mus_2 = -0.27 \text{ and } \mus_3 =
-1.22 $), with the same $\mathrm{d} \tau$, $\tilde{D}_\mathrm{p}$, and volume fraction $\phi$ as the previous
simulations (which leads to a reduced box size for the single species and species pair simulations), 5000
particles per species, and a total time $\tau_\text{total}=60$.

We provide the following movies of the Brownian Dynamics simulations:
\begin{itemize}
    \item \textbf{Movie 1}: Homogeneous mixture of species 1, which is linearly stable. See main text Figure 3(a1).
    \item \textbf{Movie 2}: Same as Movie 1, with species 2. See main text Figure 3(a2).
    \item \textbf{Movie 3}: Same as Movie 1, with species 3. See main text Figure 3(a3).
    \item \textbf{Movie 4}: Mixture of species 1 and 2, which is linearly stable.
        The particles form small, transient assemblies.
    See main text Figure 3(b1)
    \item \textbf{Movie 5}: Mixture of species 2 and 3, which is unstable,
        leading to the formation of a cluster containing both species which coexists
        with a dilute gas phase of species 2.
        See main text Figure 3(b2)
    \item \textbf{Movie 6}: Mixture of species 3 and 1, which is linearly stable.
    See main text Figure 3(b3)
    \item \textbf{Movie 7}: Mixture of all three active species,
        which undergoes self-organization, leading to the formation of a cluster
        containing all three species which coexists with a gas of species 2.
        See main text Figure 3(c).
\end{itemize}

Note that, for movies 5 and 7, and the corresponding panels in main text Figure 3, all the particle coordinates are shifted so that the centre of mass of the mixture is located at the center of the box in the final snapshot. This does not modify the result of the simulations, as they involve periodic boundary conditions, and leads to clearer figures and movies.